\documentstyle[11pt,aaspp4,flushrt,psfig]{article}

\begin{document}
\pagestyle{plain}

\title{A Photometric Investigation of the GRB970228 Afterglow and the
Associated Nebulosity}

\author{Francisco J. Castander and Donald Q. Lamb}
\affil{Department of Astronomy and Astrophysics, University of Chicago,\\
    5640 S Ellis Ave, Chicago, IL 60637}





\begin{abstract}

We carefully analyze the WFPC2 and STIS images of GRB970228. We measure
magnitudes for the GRB970228 point source component in the WFPC2 images of
$V=26.20^{+0.14}_{-0.13}$, $I_c=23.94^{+0.10}_{-0.09}$ and
$V=26.52^{+0.16}_{-0.18}$, $I_c=24.31^{+0.15}_{-0.11}$ on March 26 and
April 7, respectively; and $R_c=27.09^{+0.14}_{-0.14}$ on September 4 in the
STIS image. For the extended component, we measure magnitudes of
$R_c=25.48^{+0.22}_{-0.20}$ in the combined WFPC2 images and
$R_c=25.54^{+0.33}_{-0.22}$ in the STIS image, which are consistent with no
variation. This value is fainter than previously reported (\cite{gal98})
and modifies the previously assumed magnitudes for the optical transient
when it faded to a level where the extended source component contribution
was not negligible, alleviating the discrepancy to a power-law temporal
behavior. We also measure a color of
$V_{606}-I_{814}=-0.18^{+0.51}_{-0.61}$ for the extended source
component. Taking into account the extinction measured in this field
(\cite{CL98}), this color implies that the extended source is most likely a
galaxy with ongoing star formation.

\end{abstract}



%

\section{Introduction}

On 1997 February 28 the Gamma-Ray Burst Monitor aboard the BeppoSAX
satellite detected a gamma-ray burst (GRB) and its Wide Field Camera (WFC)
instrument imaged it. A few hours later the BeppoSAX team disseminated a
positional error circle of 3' radius to the astronomical community
(\cite{cos97a}). Eight hours after the burst, and again three days later,
the BeppoSAX Narrow Field Instrument observed this error circle. These
observations revealed a rapidly fading X-ray source at a position that was
consistent with both the WFC position for this event and with the Inter
Planetary Network (IPN) annulus calculated from the time-of-flight between
Ulysses and BeppoSAX spacecrafts (\cite{cos97b}). ASCA (\cite{yos97}) and
ROSAT (\cite{fro97},  1998b\markcite{fro98b}) observations showed that the
X-ray source continued to fade over the following two weeks, and provided a
10'' radius position for this source.

Ten days after the burst, Groot et al. (1997a)\markcite{gro97a} announced
the detection of an optical fading source ($V=21.3$) which turned out to be
the first optical counterpart detected of a GRB. Two days later Groot et
al. (1998b)\markcite{gro97b} and Metzger et al. (1997a)\markcite{met97a}
reported the presence of an extended source ($R=23.8$) at a position
consistent with that of the optical transient.  Groot et
al. (1997b)\markcite{gro97b} and van Paradjis et al. (1997)\markcite{vpa97}
claimed that the extended source was the host galaxy, but a subsequent
ground based optical observation indicated that the extended source had
faded (\cite{met97b}). Once the position of the afterglow was firmly
established, early observations were revisited and new ones taken. This new
detections provide a better sampling of the optical afterglow behavior
(\cite{klo97}; \cite{mar97}; \cite{soi97}; \cite{met97b}; \cite{ped97a};
\cite{djo97}). The HST Wide Field and Planetary Camera 2 (WFPC2) provided
another breakthrough in the follow up of the GRB970228 optical
afterglow. Observations made on March 26 revealed two components: one
consistent with being a point source and the other an extended source
(\cite{sah97a}). Further observations on April 7 showed that the point
source was fading while the extended source remained unchanged within the
observational uncertainties (\cite{sah97b}). The GRB970228 optical
afterglow was observed again with HST using the Space Telescope Imaging
Spectrograph (STIS) on September 4 (\cite{fru97}). These new images
corroborated the earlier observations: a point-like fading optical source
and an extended source, which has been claimed to be the host galaxy.


Reichart (1997)\markcite{rei97} and Wijers et al. (1997)\markcite{wij97}
discussed early observations of GRB970228 in the context of theoretical
models. They found that the gamma-ray burst afterglow behavior was
consistent with the expectations of relativistic fireball models.  Later,
Galama et al. (1998)\markcite{gal98} (revising and updating Galama et
al. (1997)\markcite{gal97}) compiled the most relevant photometric
measurements of GRB970228, converting them into a single photometric band
when necessary and subtracting the contribution of the extended source
component. They fit the optical transient temporal evolution to a power-law
with $\alpha=-1.10\pm0.04$ ($\chi^2_r=2.3$, 9 d.o.f.).

In a companion paper (\cite{CL98}; hereafter CL98), we have determined the
galactic extinction towards GRB970228, performing a careful analysis of the
publicly available observations. Here we focus on the photometric
properties of the point-like and extended components revealed in the HST
observations, and analyze their magnitudes taking into account the measured
extinction. In \S2 we detail our analysis of the WFPC2 and STIS HST
images. In \S3 we discuss our results and in \S4 we summarize our
conclusions.

\section{HST photometry}

The HST images of the GRB970228 afterglow, given their superior spatial
resolution compared to ground-based imaging, provided a startling
result. On March 26 the WFPC2 images showed two components: a point-like
and an extended source (\cite{sah97a}), whose centers were approximately
0.4{\arcsec} apart. In further observations on April 7, the point source
component was observed to fade while the magnitude of the extended
component was consistent with no variation within the measured errors
(\cite{sah97b}; \cite{sah97c}).  On September 4, the HST STIS instrument
imaged the GRB970228 field.  The point source showed further dimming while
the extended component showed no appreciable variation, after correction of
the magnitudes reported by Sahu etal. (1997c)\markcite{sah97c}
(\cite{fru97}; \cite{fru98}).

Given the outstanding importance of these measurements, we have undertaken
on a careful reanalysis of the HST photometry to discern the nature of the
GRB970228 optical counterpart in conjunction with the extinction values
measured in CL98.


\subsection{Wide Field and Planetary Camera 2 observations}

GRB970228 was observed by the HST WFC2 on March 26th and April 7th. In both
observations the optical counterpart was centered in the middle of the PC1
CCD, but there was a 2.40 deg difference in rotation angle between the
first and second observations. At both epochs, four exposures were taken in
the F606W filter and two exposures in the F814W filter, totalling 4700 and
2400 seconds, respectively (\cite{sah97c}).

In CL98 we analyzed the WFPC2 images to obtain number counts and described
the reduction process performed using standard tasks. However, in order to
accurately estimate the magnitudes and especially the errors of the objects
in the GRB position which consist of a faint point source coincident with
an extended faint surface brightness component, we developed some specific
software to carry out the photometry.

Our procedure was the following. We combined the observations in a similar
fashion as standard software does, eliminating cosmic rays by sigma
clipping with three different threshold passes cutting at 8, 5 and 3
sigma\footnote[1]{All four sets of images were clipped at these levels
except the F814W 970407 set, which presented a higher background level and
was clipped at lower sigma threshold (8,5,2.5 sigma).} and also eliminating
hot pixels masked out by the standard reduction pipeline. After combining
the images, we ended up with four values for every pixel: 1) total counts
of non-rejected pixels; 2) number of valid non-rejected pixels used; 3)
total exposure of combined pixels used; and 4) error on the total counts in
1), estimated as the square root of the counts plus the contribution of the
read-out noise.

We then calculated the centroid position of objects S1, S2 (the brightest
stars in the PC1 CCD at 2.9'' West and 16.8'' East from the GRB
respectively; see Figure~2 of CL98 for a finding chart) and the point and
extended source components of the GRB. The centroids were obtained using
the CENTER task within IRAF with the centroid algorithm. Centering in these
computed positions, the sky was estimated from circular annuli around each
object. For each object the sky was computed from 30 annuli differing in
their outer radii. A different sky value with its associated variance was
computed for pixels that had different number of rejections in the cosmic
ray and hot pixel rejection process. For the F606W filter the combination
was done with four images and typically (there is a slight difference for
each object) there were pixels with no rejections ($\sim$ 89\%), with one
rejection ($\sim$ 10\%) and two rejections ($\sim$ 1\%). For the F814W
filter the combination was done with two images and typically there were
pixels with no rejections ($\sim$ 95\%) and with one rejection ($\sim$
5\%). For every one of the 30 annuli we plotted the values of the median
and the standard deviation for each pixel with the same number of
rejections. We obtained the final sky value and its variance for each
number of pixel rejections visually from the plot. Our visual choice,
nevertheless, was similar to the median of the 30 annuli medians and the
median of the 30 annuli standard deviations. Subsequently, we calculated
aperture photometry in circular and elliptical (only for the extended
component of the GRB) apertures at radii and semimajor axes varying from 1
to 30 pixels\footnote[2]{Pixel scale: $0.0455''$/pixel} around each
object. We added the pixels values (and fractional pixel values at the
edges) within the apertures, subtracting the sky values for the appropriate
number of pixel rejections, taking into account the error of both counts
and sky, and normalizing to a common exposure. Each aperture had therefore
a properly computed count flux and error.

\begin{figure}[t]
\columnwidth10cm
\plotone{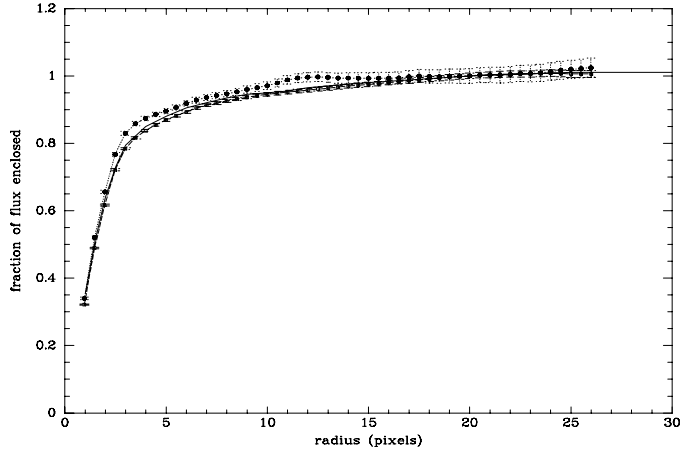}
\vspace{-0.0cm}
\caption[]
{Point spread function for star S1 (dotted line and solid circles), S2
(dashed line and open squares) in the combined F606W WFPC2 images taken on
March 26 and that defined by Holtzman et al. (1995a)\markcite{hol95a} (solid line).  }
\label{fig1}
\end{figure}

We used stars S1 and S2 to define the point spread function
(PSF). Figure~\ref{fig1} shows the normalized counts for both stars at all
apertures used together with the approximate encircled energy curve for the
PC1 given by Holtzman et al. (1995a)\markcite{hol95a} that serves as
comparison.

\begin{figure}[t]
\plotone{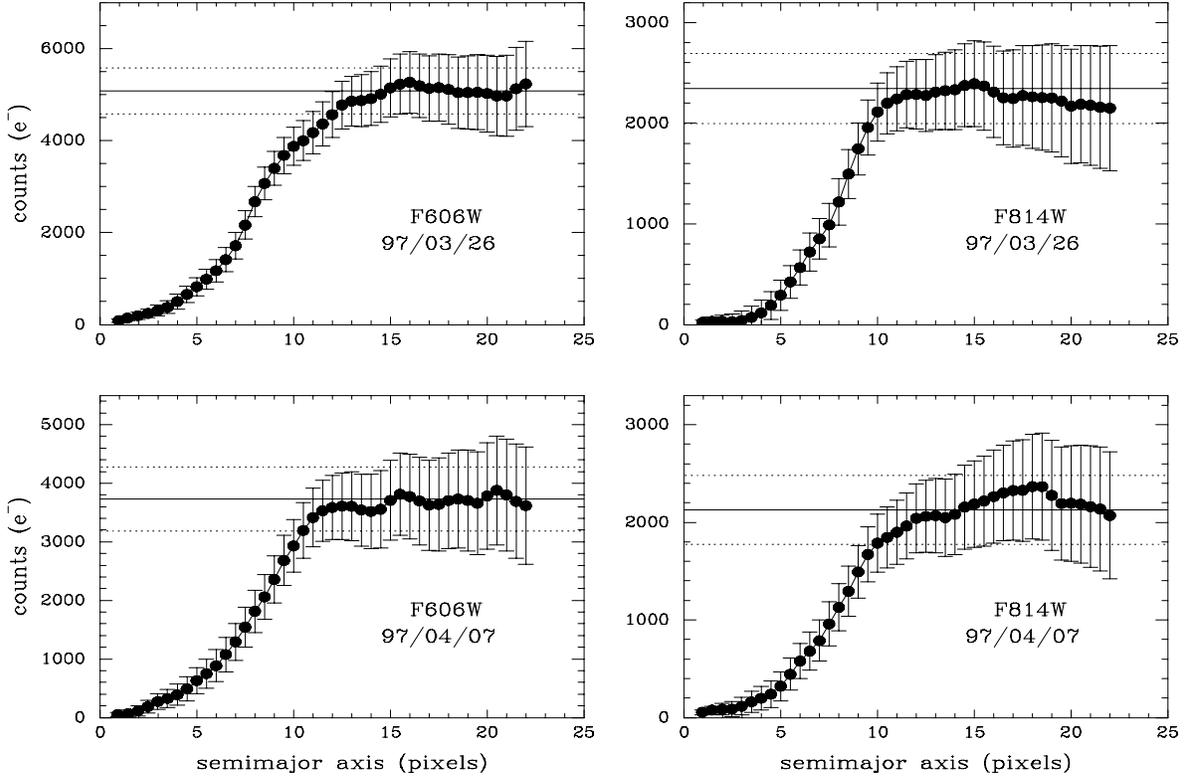}
\vspace{-0.0cm}
\caption[]
{Enclosed counts as a function of semimajor axis for the extended+point
components in the WFPC2 images. The solid line illustrates the value
adopted and the dashed lines its error.}
\label{fig2}
\end{figure}

Table~\ref{tbl3} summarizes our photometric measurements. We first analyzed
the point source component. We measured its magnitude minimizing the
$\chi^2$ of the fit of the inner radial bin counts of the point source to
the same bins for stars S1, S2 and the PC1 PSF, where the normalization of
the stars is the free parameter being fit. We used five bins: the first one
corresponds to the counts in a 1 pixel radius circle and the other four, to
the counts in successive concentric circular annuli of 0.5 pixels
width. That is, we fit the inner 3.0 radius circle counts to three model
PSFs. The extended component was measured in circular and elliptical
apertures, beginning at the estimated centroid and extended out to 30
pixels radius/semimajor axis. Then we plotted the counts enclosed in every
aperture as a function of aperture radius/semimajor axis and assigned the
counts of the extended and pioint components to the value at which the
counts remained approximately unchanged with increasing radius. We
estimated the error as the error in the aperture where the counts started
to flatten, $a\simeq12$ pixels (see Figure~\ref{fig2}). We found that the
exact choice of the centroid and ellipticity did not affect our
measurements. These measured counts and errors correspond to both the point
and extended components, as both lie within the apertures where the counts
flatten. Then we calculated the extended component magnitude by subtracting
the point source contribution to the total counts. Once we had this first
estimation of the extended source, we then estimated the extended source
contribution to the pixels where the point component was measured and redid
the $\chi^2$ fit to the point source including this contribution. Having
obtained this new value and the previously measured point+extended counts,
we again calculated the extended component magnitude and its error.

\subsection{Space Telescope Imaging Spectrograph observations}

On September 4 the STIS instrument onboard HST imaged the GRB970228 from
4.6601 to 4.7657 UT. Two exposures of 575s each were taken in the clear
aperture (50CCD) mode at each of four dithered positions for a total
exposure time of 4600s. The dithered positions were at 0.125'' SW, 0.125''
SE and 0.250'' S offsets from the first image for the second, third and
fourth images respectively. We retrieved the pipeline-processed images from
the HST archive once they became public almost two months after being
taken. We reduced them in a similar fashion as was done with the WFPC2
images, although in this case we had to shift the images to correct for the
dithering, which required some rebinning, as the offsets were not done at
integer pixels for all four images. We again ended up with an image having
four values per pixel: 1) total number of non-rejected counts; 2) number of
valid non-rejected pixels used; 3) total non-rejected exposure; and 4) error
on the total counts.

\begin{figure}[t]
\columnwidth10cm
\plotone{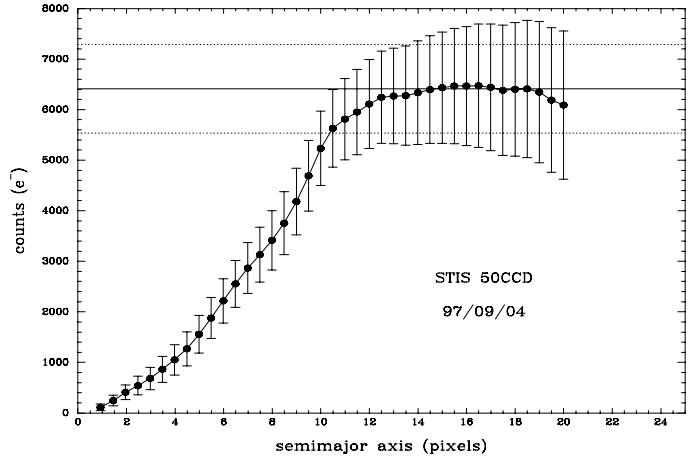}
\vspace{-0.0cm}
\caption[]
{Enclosed counts as a function of semimajor axis for the extended+point
components in the STIS image. The solid line illustrates the value adopted
and the dashed lines its error.}
\label{fig3}
\end{figure}

We then proceeded as we had with the WFPC2 images, obtaining the sky value
and variance for every object, and the number of rejected pixels. In the
STIS images typically 96\% of the pixels had no rejections, 4\% one
rejection and less then 0.1\% had two rejections. Table~\ref{tbl4}
summarizes our photometry. Magnitudes were computed as for the WFPC2
images.

\section{Discussion}


We have undertaken an effort to determine the photometry of the optical and
extended components of the GRB970228 optical counterpart, as observed by
the HST WFPC2 and STIS instruments. We have tried to perform the best
magnitude determination with a rigorous treatment of the errors. As
described in the previous section, we started by combining the original
pipeline processed images, retaining the rejection information to propagate
it into our error analysis. Our subsequent sky subtraction was based on a
multi-aperture estimation that takes into account the number of rejected
pixels in the combination process. Although the sky varies depending on the
exact area used to estimate it, our multi-aperture estimation reduces
problems that could arise from selecting an area with deviant values. Our
error also takes into account the variance of the sky estimation due to the
selected area. Although the fraction of rejected pixels per unit area is
relatively stable throughout all regions of the images studied, we also
apply a different sky correction depending on the number of rejected
pixels.

We have chosen to determine the magnitude of the point source component by
fitting the inner counts radial profile to the PSF. We use the radial
profile within the inner 3.0 pixels (0.136'') for the WFPC2 images and the
inner 2.0 pixels (0.1'') for the STIS images, taking into account in a
second phase the contribution of the extended component as well. The radius
used in the STIS analysis was smaller in an attempt to reduce the
uncertainties of the contribution of the extended source, as the point
source is significantly fainter in this observation. Once the fit is done,
we extrapolate the PSF profile to obtain the total counts (see previous
section for further details). We believe that this is the best procedure to
obtain the point component magnitude and its error. The method should give
results that are, however, similar to obtaining aperture photometry and
appling a correction factor for that aperture to get the total
magnitudes. Our method is essentially the same, but performed for several
radii. We use the two brightest stars in the field, S1 and S2, and the
Holtzman et al. (1995a)\markcite{hol95a} and Robinson et
al. (1997)\markcite{rob97} determination of the PSF for WFPC2 and STIS,
respectively, as the model PSF to which the point component data was fit.
Table~\ref{tbl5} provides the reduced $\chi^2/\nu$ values for the different
filters and epochs of the observations. Fits are acceptable in all cases
except for the April F606W and the STIS images. In both cases the radial
count distribution profile is more extended than that of the stars and
model PSF. Our choice of measuring method by profile fitting is not optimal
in these two cases, but is more accurate than single aperture
photometry. The outermost radius we use in the fits encloses approximately
80\% and 65\% of the counts (WFPC2 and STIS images respectively), so our
extrapolation of the profile only contributes at most $\sim$ 35\% of the
total counts. If we change the outermost radius in the fits our
extrapolated total counts do not vary appreciably. Only in the two cases
where the PSF fit is poor is the variation somewhat larger. Finaly, we
would like to emphasize that we include an estimated contribution of the
extended component in the pixels measured when performing the fit.

We estimate the total counts of the extended+point components by elliptical
aperture photometry (see previous section). We increase the semimajor axis
until the counts flatten with radius. Due to different centroiding, the
counts do not flatten at the same semimajor axis value in all cases. In
Figure~\ref{fig2} we can see how the number of counts levels off and
remains approximately constant with increasing semimajor axis for the F606W
filter, providing a relatively safe count estimation. For the F814W filter,
however, the number of counts still varies somewhat at large radii. In this
case the image combination was performed only with two images and the
cosmic ray/hot pixel removal is more difficult. For example, if we had
chosen a lower threshold rejection in the combination process, the increase
in the number of counts in the F814W 97/04/07 image at semimajor axis
values from 15 to 18 pixels disappears, as this feature is due to a count
excess present in only one of the images that was not rejected properly
with the threshold used.

With the measured counts and exposure times we compute ST
magnitudes\footnote[3]{All magnitudes that we quote for the GRB97028 point
and extended components are in the ST system for the WFPC2 filters and in
the Vega system for standard Johnson-Cousins filters} for the point and
extended components using the header keywords PHOTFLAM and PHOTZPT in the
WFPC2 images (Table~\ref{tbl3}). For the STIS image we present count rates
(Table~\ref{tbl4}), since the STIS ST magnitude in open mode is not very
useful, given that an assumed spectrum is required to convert STIS 50CCD
count rates to other broad band filter magnitudes. In order to compare the
HST observations with other ground-based magnitude measurements, we convert
our values to Johnson-Cousins magnitudes. Instead of using color
transformation relations (\cite{hol95b}), we fit a spectrum to the observed
colors, convolving the spectrum with the filter responses. We then compute
the expected magnitude in a particular filter with the spectral fit.

We fit a power-law spectrum for the point source and a galaxy spectrum for
the extended component. In both cases we assume a foreground $R_V=3.1$
extinction law of $A_V=1.19$ (CL98) and no intrinsic reddening at the
source. Obviously, the reddening assumed for the point source component is
irrelevant, since the extinction law behaves approximately as a power law
in the wavelength range used and the only effect caused is a change in the
power-law slope of the fit. The point source spectrum is fit at both epochs
for the WFPC2 images. The best-fit power-law indexes are: 1.17 and 1.06 for
March 26 and April 7, respectively. To compute standard filter magnitudes
for the point source from the STIS number counts we assume an $A_V=1.19$
extincted power-law spectrum that gives the mean of the color between the
two WFPC2 observations, $V_{606}-I_{814}=0.70$. The errors in the
magnitudes come from varying the power-law indeces to allow for all
possible values of the observed $V_{606}-I_{814}$ color in the WFPC2. We
assumed that there was no color evolution for the point source component
from the March 26 and April 7 WFPC2 observations to the September 4 STIS
observations.

We have combined the two WFPC2 observations to provide the best magnitudes
and errors for the extended component. We add the measured counts,
weighting them by their relative errors. With them, we compute the
resulting magnitudes and errors. We obtain $V_{606} =
26.03^{+0.19}_{-0.16}$ and $I_{814}=26.21^{+0.45}_{-0.33}$. For the
extended source we assume a galaxy spectrum to convert from the combined
HST measurements to standard filters magnitudes. There are several galaxy
templates at different redshifts that reproduce the observed colors. We
choose an $A_V=1.19$ extincted spectrum of a starburst galaxy with constant
star formation observed at $z=1.46$, 1 Gyr after the onset of the burst
(see below) to convert from the best-fitting magnitudes in the WFPC2
filters to Johnson--Cousins'. To allow for the possible spectral
differences allowed by the observed color errors, we compute the color
transformation for a variety of synthetic galaxy spectra that bracket the
measured error in the $V_{606}-I_{814}$ color. Therefore, the errors in the
standard filters include the measured color error plus the contribution due
to the different spectral energy distributions (SED) allowed. Our search of
SEDs compatible with the observed colors has been extensive but by no means
complete. Nevertheless, we believe that the estimated contribution to the
error due to the allowed SEDs is appropriate.  In the STIS case, we proceed
in a similar fashion to compute magnitudes from count rates. As for the
WFPC2 case, the best-fitting value is obtained with a spectrum of a 1 Gyr
old burst of star formation observed at $z=1.46$; and its error, from the
combination of the contribution due to the error in the measured count rate
and the contribution due to the allowed SEDs as determined by the measured
colors in the WFPC2 observations.  Tables~\ref{tbl3} and \ref{tbl4} present
our measured magnitudes.

Once the magnitudes of point and extended sources have been measured, what
can we learn about them? For the point-source component, our magnitudes are
consistent with those measured by Sahu et al. (1997c)\markcite{sah97c} in
the $V$ band but are 0.3 magnitudes brighter in $I_c$. Overall, our
computed color ($V-I_c=2.23$) is redder than that of Sahu et al
($V-I_c=1.85$). We have assumed a power-law spectrum to convert between ST
and standard filter magnitudes (see above). Our magnitudes do not
appreciably change if we were to use instead the photometric transformation
of Holtzman et al. (1995b)\markcite{hol95b}. Our $R_c$ magnitudes are,
nevertheless, almost identical to those computed by Galama et
al. (1997)\markcite{gal97} from Sahu et al's data. We are unable to comment
on the discrepancy of the $I_c$ magnitudes as Sahu et
al. (1997c)\markcite{sah97c} do not provide enough details for a
comparison. In the STIS image we measure a count rate of
$0.232^{+0.023}_{-0.022}$ for the point source. Our estimated $V$ and $R_c$
magnitudes are consistent within the errors with those measured by Fruchter
et al. (1998)\markcite{fru98} and Galama et
al. (1998)\markcite{gal98}. Fruchter et al. (1998)\markcite{fru98} measure
$V=28.0\pm0.25$ while we get $V=28.10^{+0.24}_{-0.23}$. The 0.1 mag
difference in our best values could be entirely due to the spectrum assumed
to convert from STIS counts to magnitudes. In the $R_c$ filter, our
measurement, $R_c=27.09^{+0.24}_{-0.23}$, is more discordant to that of
Galama et al. (1998)\markcite{gal98}, $R_c=27.25\pm0.27$. In this case, the
conversion between STIS counts and $R_c$ magnitudes is less dependent on
the assumed spectrum (as can be seen in Table~\ref{tbl4}, where the errors
bars are smaller for this filter) and the 0.16 mags difference in our best
values is probably due to different measured count rates. There are
indications that the OT is becoming redder in its evolution (e.g.,
\cite{gal98}). The spectrum assumed to convert count rates to magnitudes
has not taken into account this possibility. However, the effect should be
small and should not affect our $R_c$ magnitude appreciably.

\begin{figure}[t]
\plotone{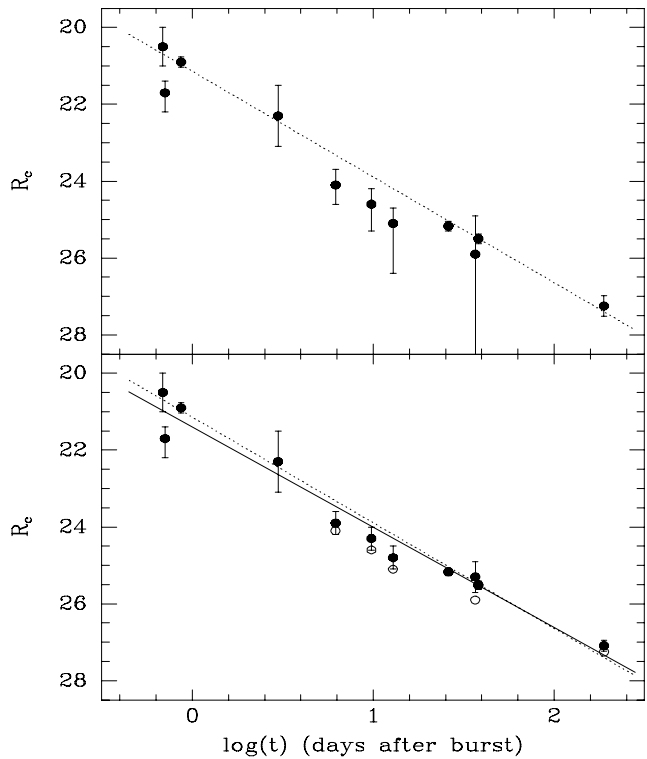}
\caption[]
{The $R_c$ light curve of GRB970228. Top Panel: light curve by Galama et al. (1998)\markcite{gal98}
with their power-law fit ($\alpha =-1.10$, dotted line). Bottom panel:
light curve with OT magnitudes revised with our value of the extended
component magnitude with our best power-law fit ($\alpha =-1.04$, solid
line). For comparison, the Galama et al's points are plotted as open
circles, their power-law fit as dotted lines.}
\label{fig4}
\end{figure}

For the extended component, our measured magnitudes for the WFPC2 and STIS
images agree remarkably well. To convert from STIS count rate to
magnitudes, we have assumed a spectrum that fits the WFPC2 colors. That is
the reason why the colors for the extended component are the same in the
WFPC2 and STIS images (Tables~\ref{tbl3} and \ref{tbl4}). Compared to
previous work, our $V$ magnitudes agree within the errors with those of
Fruchter et al. (1997)\markcite{fru97} and Fruchter et
al. (1998)\markcite{fru98} ($V=25.6\pm0.25$ and $V=25.7\pm0.25$ for WFPC2
and STIS respectively). However, we disagree with the $R_c=25.0\pm0.3$
magnitude quoted by Galama et al. (1998)\markcite{gal98}. Combining the
WFPC2 and STIS magnitudes we obtain $R_c=25.51^{+0.18}_{-0.15}$. This
magnitude would imply a slightly different temporal behavior for the OT
than that reported by Galama and collaborators, because they overcorrected
for the contribution of the extended component when the OT faded to fluxes
comparable to that of the extended component. Figure~\ref{fig4} shows the
$R_c$ light curve of Galama et al. (1998)\markcite{gal98} (top panel) along
with our revised light curve (bottom panel). In the bottom panel we also
plot the Galama et al's points as open circles to compare the difference
that the revised extended component magnitude has on the OT magnitudes. We
plot both curves in order to compare error bars as well. The new revised OT
magnitudes produce a better fit to a power-law behavior than before at late
epochs (after the first week). If we consider only the last 7 photometric
points, that is, the evolution after March 06, we obtain a power-law
temporal decay of $\alpha=-0.86\pm0.06$ ($\chi^2/\nu=0.27$, 5 d.o.f). The
same points with the Galama and collaborators values would give
$\alpha=-0.82\pm0.10$ ($\chi^2/\nu=0.71$, 5 d.o.f). However, considering
all the photometric points we obtain $\alpha=-1.04\pm0.03$
($\chi^2/\nu=2.49$, 9 d.o.f.; note the overly small errors in the slope due
to the poor $\chi^2$ fit) .vs. the $\alpha=-1.10\pm0.04$ ($\chi^2/\nu=2.3$,
9 d.o.f) of Galama et al. Therefore, it appears that the temporal decay is
slowing down or that the early behavior differs from the extrapolation of
the later temporal behavior to earlier times, although caution is necessary
when interpreting the first photometric measurements, since for example,
the second point disagrees with the first and third. We believe this
explanation (steady power-law decay at {\it late} times and uncertain
photometric measurements at {\it early} times) is more plausible than a
strong deviation from the power-law behavior at ``intermediate'' times, as
suggested by Galama et al. (1998)\markcite{gal98}. The apparent shallower
power-law behavior could be also a manifestation of the OT becoming redder.

\begin{figure}[t]
\vspace{-0.8cm}
\plotone{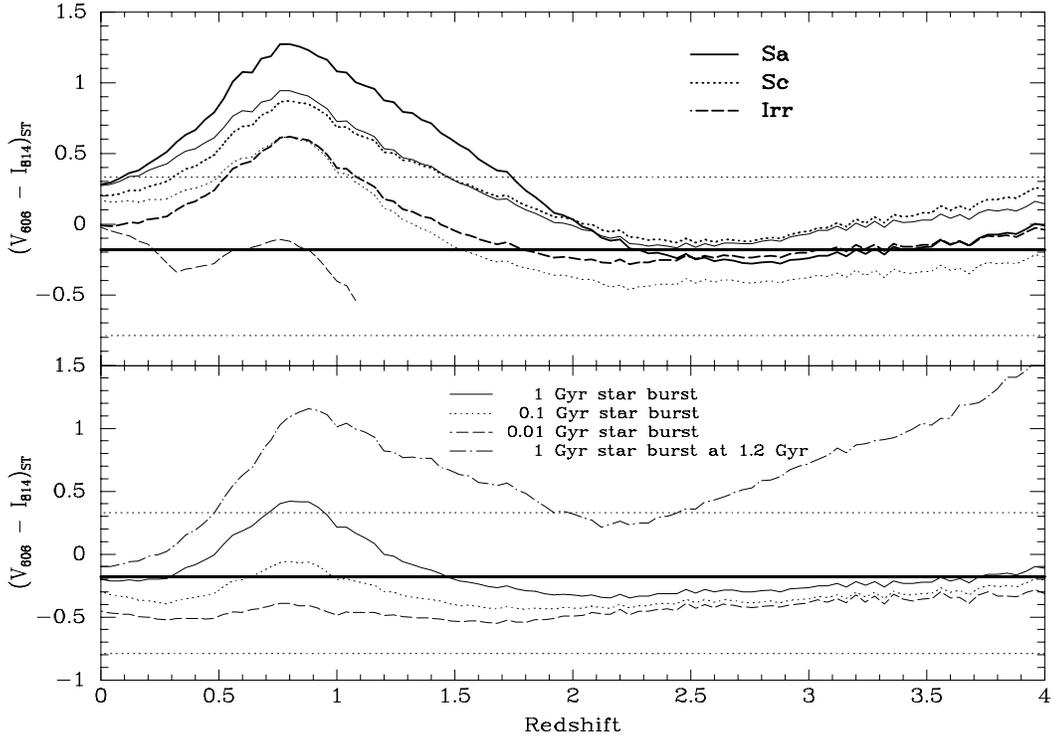}
\caption[]
{WFPC2 colors for synthesized Spectral Energy Distributions (SEDs)
.vs. redshift. The top panels shows the expected colors for spirals and
irregular galaxies including a local extinction of $A_V=1.19$. The thick
lines include a K-correction, the thin lines include evolutionary plus
K-corrections. The thick horizontal line is the best value observed color
and the horizontal dotted lines its $1\sigma$ error. The bottom panel shows
models of 1 Gyr (solid line), 100 Myr (dotted line) and 10 Myr (dashed
line) starbursts observed at redshift $z$. The dot-dashed line represent a 1
Gyr startburst observed at redshift $Z$ 200 Myr after the termination of
the star formation. The truncation of the star formation with the
disappearence of the most massive stars produces redder colors incompatible
with the observations.}
\label{fig11}
\end{figure}

What can we learn from the HST observations of the extended component?
Several authors (e.g, \cite{sah97c}; \cite{liv98}; \cite{fru98}) have
previously discussed the properties of the extended component. Fruchter et
al. (1998)\markcite{fru98} find that the size of the extended component is
consistent with the sizes of galaxies of comparable magnitude in the
HDF. Based on this and the surface number density of galaxies of similar
surface brightness, they conclude that the extended source is most likely a
galaxy at moderate redshift, and is almost certainly the host galaxy of
GRB970228. Here, we concentrate on the observed colors, taking into account
the measured extinction, in order to constraint the nature of the likely
galaxy. For that purpose, we will compare them to the colors obtained from
synthetic galaxy spectra computed with PEGASE (\cite{FRV97}), a galaxy
synthesis code that reproduces the observed colors of nearby galaxies.

As mentioned before, for the extended component we measure a color of
$(V_{606}-I_{814})_{ST} = -0.18^{+0.51}_{-0.61}$ in the WFPC2 images and a
count-rate of $1.161\pm0.192$ in the STIS image, and we obtain a local
galactic extinction of $A_V=1.19^{+0.10}_{-0.17}$ (CL98). We prefer to use
the measured color in the WFPC2 filters and the count-rate in the STIS
image to avoid introducing errors in transformations to other
filters. Figure~\ref{fig11} (top panel) presents the expected colors,
including a local $R_V=3.1$ extinction law (\cite{car87}; \cite{odo94}) of
$A_V=1.19$, for three different spectral energy distributions corresponding
to Sa (solid line), Sc (dotted line) and Irregular (dashed line) galaxies
as a function of redshift, together with the observed color (solid line:
best value, dotted lines: $1\sigma$ errors). We have obtained the
$V_{606}-I_{814}$ color in the redshift range $0<z<4$ for each synthetic
galaxy spectra in two ways. First, we have redshifted the spectra, which
corresponds to applying a K-correction (thick lines). We have also
redshifted the spectra evolving them back in time with a star forming
history prescription that reproduces the observed colors at $z=0$, which
correspond to an evolutionary plus K-correction (thin lines). From
Figure~\ref{fig11} we can check that an Sa galaxy is only consistent with
our data if it is at $z\gtrsim1.5$; an Sc, if $z \lesssim 0.4$ or $z
\gtrsim1.2$ and the constraint on irregulars depend strongly on their
assumed formation time. In order to clarify what are the galaxy stellar
contents allowed by the observed colors we plot in the bottom panel of
Figure~\ref{fig11} the prediction of a 1 Gyr (solid line), a 100 Myr
(dotted line) and a 10 Myr (dashed line) burst of star formation observed
at redshift $z$ at the termination of the burst. That is, the burst starts
1 Gyr, 100 Myr and 10 Myr, respectively, before it is observed at redshift
$z$. We also show a 1 Gyr burst model that is observed at redshift $z$ 200
Myr after the end of the burst (dot-dashed line) to understand the effect
of the truncation of star formation. As Figure~\ref{fig11} (bottom panel)
shows, on-going bursts of star formation of duration shorter than 1 Gyr
produce acceptable colors; longer duration bursts begin to be unacceptable
at around redshift $z\sim0.8$. This is because, for longer duration bursts,
the contribution to the total light of evolved stars increases with respect
to that of the most massive stars (O and B) that are continuosly
formed. When the A stars start to dominate the total emitted light there is
a significant flux decrement in the region of the most energetic Balmer
lines ($3645-4000$ {\AA}). This decrement passes between the F606W and
F814W filters around $z\sim0.8$, explaining the redder $V_{606}-I_{814}$
color in the bottom panel of Figure~\ref{fig11}. Older stellar populations
would produce even redder colors as late type stars contribute more
and more to the integrated light (see top panel). Another important
conclusion comes from the truncated star formation model (dotted-dashed
line). In this case, after 200 Myr without star formation, all hot stars (O
and B) have evolved off the main sequence and there is not enough
ultraviolet flux to explain the observed colors if the redshift is
$z\gtrsim0.5$.

Summarizing, the observed $V_{606}-I_{814}$ color requires the presence of
hot stars dominating the ultraviolet flux if the galaxy is at
$z\gtrsim0.5$, implying on-going star formation. If the galaxy is composed
of stellar populations similar to present day spirals or their likely
progenitors, then its redshift is $z\gtrsim1.2$, unless it is a dwarf at
least 4 magnitudes fainter\footnote[4]{Taking into account the measured
local extinction} than an L$^{\star}$ (and semimajor axis $\lesssim 2.5
h^{-1}$ kpc), in which case it could also be at $z\lesssim0.5$, but in this
case it must be unusually faint for its size.

\begin{figure}[t]
\plotone{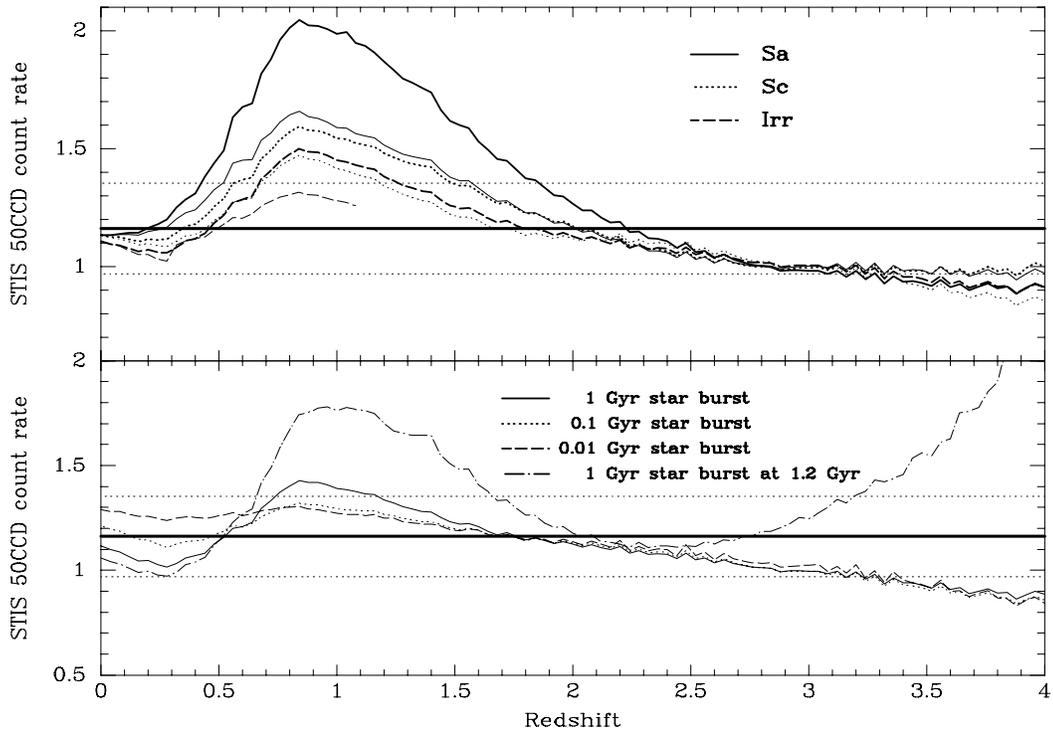}
\caption[] 
{STIS count-rates for synthesized SEDs .vs. redshift. The models presented
are the same as in Figure~\ref{fig11} but are normalized to a
$(V_{606})_{ST}=26.03$ magnitude as observed in the combined WFPC2 images.}
\label{fig12}
\end{figure}

In Figure~\ref{fig12} we present a similar plot, using the same synthetic
spectra, but now we predict the STIS 50CCD count rate instead of the
$V_{606}-I_{814}$ color. In this case, we normalize the synthetic spectra
to give the observed $V_{606}=26.03$ magnitude in the combined WFPC2
image. This only differs by 0.05 magnitudes from the $V_{606}$ magnitudes
we obtain from the measured STIS 50CCD count-rate assuming a galaxy
spectrum that reproduces the color observed in the WFPC2 images (see
above and Table~\ref{tbl4}). In fact, in this way, given the wide
response of the 500CCD STIS observing mode ($\sim2000-10000$ {\AA}), we are
measuring the change in count-rate due to the change of spectral shape (or
color) as we fix the normalization at $V_{606}$. The conclusions derived
from the STIS count-rate (Figure~\ref{fig12}) are almost the same as those
derived from the WFPC2 colors. For the STIS count-rate, however, the
constraint on the allowed redshift ranges for {\it typical} spirals are
even tighter at high redshift, $z\gtrsim1.3$. Even bursts of star formation
start to have a problem  explaining the measured count-rate at $z\sim1$. We,
therefore, conclude in view of the measured magnitudes and count-rate that
the extended object is most likely a galaxy undergoing star formation at a
redshift possibly higher than $z\sim1.3$.

\begin{figure}[t]
\plotone{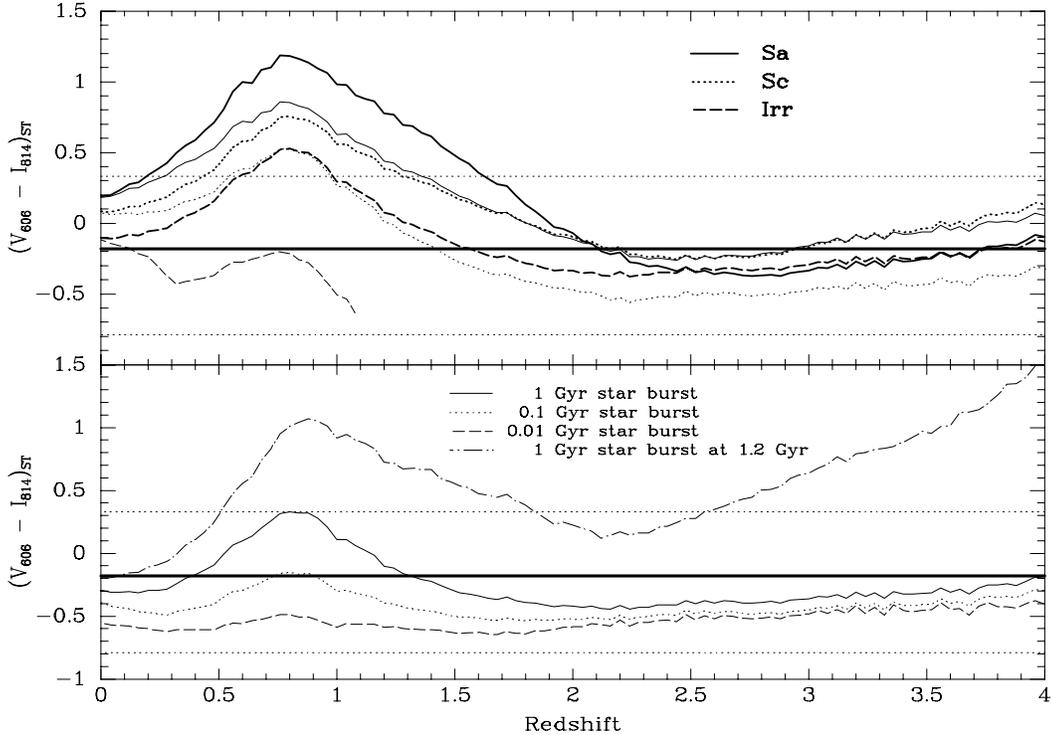}
\caption[]
{Same as figue~\ref{fig11}, but for a local extinction of $A_V=0.86$.}
\label{fig13}
\end{figure}

\begin{figure}[t]
\plotone{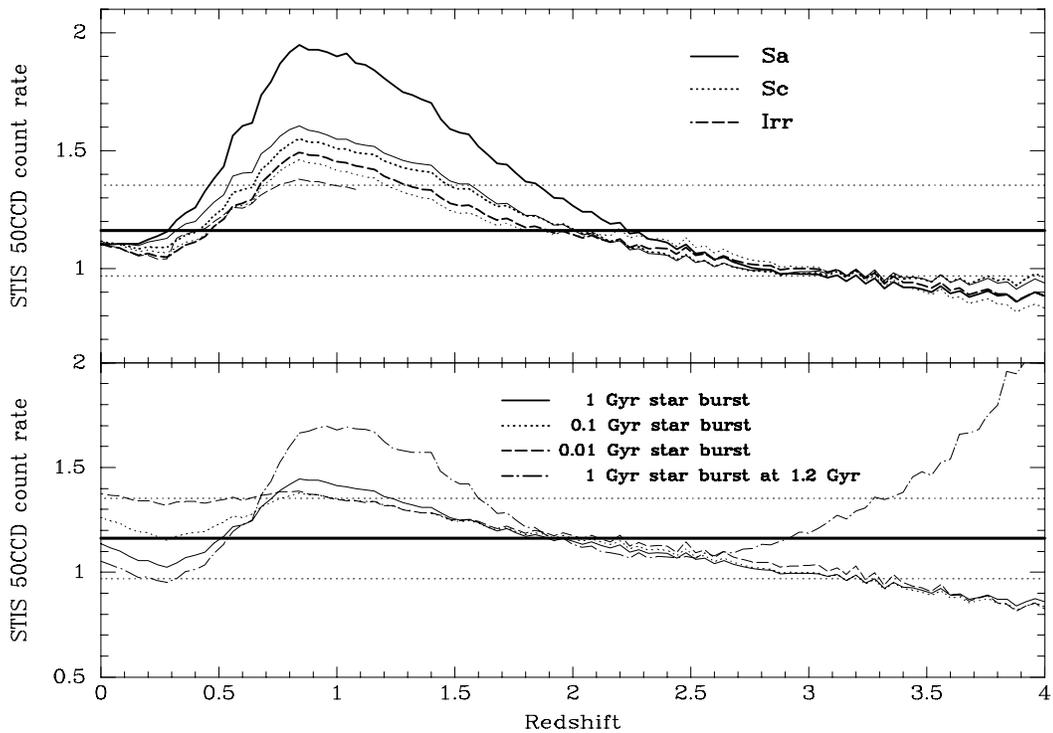}
\caption[]
{Same as figue~\ref{fig12}, but for a local extinction of $A_V=0.86$.}
\label{fig14}
\end{figure}

One possible source of uncertainty to our conclusion is the assumed value
of the local galactic extinction. We have used the best value obtained in
CL98 ($A_V=1.19^{+0.10}_{-0.17}$), which is the result of the combination
of several estimates. This value is pushed high by the extinction measured
from the spectra of stars in the vicibity of GRB970228. If in CL98 we had
not used the extinction estimate from the stellar spectra, we would have
obtained a weighted best value of $A_V=0.86\pm0.11$ (see CL98's
Table~2). Figures~\ref{fig13} and~\ref{fig14} are the equivalent to
Figures~\ref{fig11} and~\ref{fig12} showing the same synthesized spectra,
but now for $A_V=0.86\pm0.11$. As can be seen, our conclusions remain
essentially unchanged.

In our analyses of magnitudes and colors of the extended component so far,
we have assumed no intrinsic extinction at the source. If the extended
component is a galaxy with ongoing star formation, it is likely that it
will contain dust and therefore absorb part of the optical radiation
emitted (e.g., \cite{cal94}). Should any intrinsic extinction be present,
our conclusion of a galaxy with ongoing star formation would be
strengthened.

Tonry et al. (1997)\markcite{ton97} and Kulkarni et
al. (1997)\markcite{kul97} have tried to obtain the spectrum of the
GRB970228 afterglow and its associated nebulosity. When the first set of
observations were taken, the magnitude of the OT was similar to that of the
nebulosity, while for the second set the former was significantly
fainter. Neither observation revealed any obvious emission lines. If the
extended source is a galaxy with ongoing star formation, strong emission
lines are expected. The lack of observed [OII] and Ly${\alpha}$ emission
lines suggests that the galaxy lies at a redshift between $1.5\lesssim z
\lesssim 2.6$ for the spectral coverageof the observations, in which case
the colors and size observed for the extended component would be most
plausibly interpreted as those of a star forming galaxy.

\section{Conclusions}


We reanalyze the images of GRB970228 field taken with the HST WFPC2 and
STIS instruments. Tables~\ref{tbl3} and \ref{tbl4} summarize our
photometric measurements. As previously reported, we find that the point
source component has faded between the time elapsed by these observations
while the extended component shows no significant variation. Although a
possible reddening of the GRB970228 optical transient colors has been
reported (\cite{gal98}), no significant color change is detected for the
point source component between the two WFPC2 images.

Analyzing the two WFPC2 images together, we find for the extended source
$V_{606}=26.03^{+0.19}_{-0.16}$ and $I_{814}=26.21^{+0.45}_{-0.32}$, which
we transform (see above) into $R_c=25.48^{+0.22}_{-0.20}$. This magnitude
is significantly fainter than that reported by Galama et
al. (1998)\markcite{gal98}. Ground-based optical observations of the point
optical transient suffered some contamination from the extended
component. This contamination was negligible during the first days after
the onset of the burst, but substantial after the first week. Therefore,
our measured $R_c$ magnitude implies that the total OT+extended magnitudes
compiled by Galama et al. (1998)\markcite{gal98} were overcorrected for the
extended source by these authors in obtaining the magnitude of the
OT. Using our $R_c$ magnitude, we find that the temporal behavior behaves
of the OT is consistent with a power-law decline after the sixth day of the
burst, with a temporal decay of $\alpha=-0.86\pm0.06$, and shows no
significant deviations. The extrapolation to early epochs (before the end
of the first day after the burst) would agree with the observed magnitude
at Feb 28.83 UT (\cite{gua97}) but disagree with the observations at Feb
28.81 UT (\cite{ped97b}) and Feb 28.99 UT (\cite{gal97}). If we choose to
fit all of the photometric points, we obtain a temporal decay slope of
$\alpha=-1.04\pm0.03$, but the data is poorly fit, in part due to the
inconsistency amongst the early measurements (see
Figure~\ref{fig4}). Given the small error bar in the WHT Feb 28.99 UT
measurement and ignoring any possible zero-point error (the observation was
not carried out in the R-band filter), the temporal behavior of the OT
appears to have changed from a steep early decline to a shallower power-law
decay after the first week. Some of this spectral slope change could be do
to intrinsic reddening of the source. However, the light curve is open to
multiple interpretations because the number of observations is sparse.

The WFPC2 magnitudes and STIS count-rate of the extended source component
are consistent with no variation, within the errors. Several considerations
indicate that this object is most likely a galaxy and possibly the host of
GRB970228. If this object were a galaxy, its color,
$V_{606}-I_{814}=-0.18^{+0.51}_{-0.61}$, is remarkably blue when we take
into consideration the measured extinction. Using synthetic spectra we
conclude that the observed emission from this object is dominated by hot
stars and therefore is most likely a starburst galaxy or a spiral at high
redshift. We tentatively show that its redshift is likely to be
$z\gtrsim1.2$. Further considerations of failed detection of emission lines
would place this object between $1.5\lesssim z \lesssim 2.6$.

\acknowledgments

We acknowledge valuable discussions with Daniel Reichart, Mark Metzger,
Andrew Fruchter, Carlo Graziani, Jean Quashnock, Cole Miller and Dave
Cole. Part of this work is based on NASA/ESA Hubble Space Telescope
archival data retrieved from the archive maintained at STSci. We
acknowledge support from NASA grants NAGW-4690, NAG 5-1454, and NAG 5-4406.


\clearpage

%

\begin{deluxetable}{ccccc}
\tablecaption{WFPC2 Magnitudes of the GRB point and extended
components.\label{tbl3}} 
\tablewidth{12cm} 
\tablehead{ \colhead{} & \colhead{Point} & \colhead{Date} & \colhead{Extended} & \colhead{Date}} 
\startdata
$V_{606}$ & $26.09^{+0.08}_{-0.08}$ & 97/03/26 & $25.91^{+0.23}_{-0.19}$ & 97/03/26 \nl 
$V_{606}$ & $26.42^{+0.09}_{-0.10}$ & 97/03/26 & $26.25^{+0.36}_{-0.27}$ & 97/04/07 \nl 
$I_{814}$ & $25.37^{+0.09}_{-0.09}$ & 97/04/07 & $26.50^{+1.06}_{-0.52}$ & 97/03/26 \nl 
$I_{814}$ & $25.74^{+0.14}_{-0.11}$ & 97/04/07 & $26.05^{+0.57}_{-0.37}$ & 97/04/07 \nl 
\tableline
$V$       & $26.20^{+0.14}_{-0.13}$ & 97/03/26 & $25.85^{+0.26}_{-0.23}$ & Combined\nl 
$R_c$     & $25.17^{+0.09}_{-0.08}$ & 97/03/26 & $25.48^{+0.22}_{-0.20}$ & Combined\nl 
$I_c$     & $23.94^{+0.10}_{-0.09}$ & 97/03/26 & $24.84^{+0.46}_{-0.42}$ & Combined\nl 
$V$       & $26.52^{+0.16}_{-0.18}$ & 97/04/07 &  & \nl 
$R_c$     & $25.52^{+0.11}_{-0.11}$ & 97/04/07 &  & \nl 
$I_c$     & $24.31^{+0.15}_{-0.11}$ & 97/04/07 &  & \nl
\enddata
\end{deluxetable}

\begin{deluxetable}{ccc}
\tablecaption{STIS Magnitudes of the GRB point and extended
components.\label{tbl4}} 
\tablewidth{10cm} 
\tablehead{ \colhead{} & \colhead{Point} & \colhead{Extended}} 
\startdata
Count rate & $0.232^{+0.023}_{-0.022}$ & $1.161^{+0.192}_{-0.192}$ \nl
\tableline 
$V_{606}$ & $27.99^{+0.20}_{-0.18}$ & $26.09^{+0.35}_{-0.24}$ \nl 
$I_{814}$ & $27.29^{+0.19}_{-0.16}$ & $26.27^{+0.85}_{-0.53}$ \nl 
$V$       & $28.10^{+0.24}_{-0.23}$ & $25.92^{+0.40}_{-0.27}$ \nl 
$R_c$     & $27.09^{+0.14}_{-0.14}$ & $25.54^{+0.33}_{-0.22}$ \nl 
$I_c$     & $25.87^{+0.19}_{-0.16}$ & $24.90^{+0.86}_{-0.63}$ \nl 
\enddata
\end{deluxetable}

\begin{deluxetable}{cccccc}
\tablecaption{GRB afterglow point source component fit to PSF.\label{tbl5}} 
\tablehead{ \colhead{} & \colhead{\small F606W{\footnotesize (03/26)}} & \colhead{\small F606W{\footnotesize (04/07)}} & \colhead{\small F814W{\footnotesize (03/26)}} & \colhead{\small F814W{\footnotesize (04/07)}} & \colhead{\small STIS{\footnotesize (09/04)}}} 
\startdata
$\chi^2/\nu$ & 0.73 & 5.71 & 0.21 & 0.38 & 5.77 \nl
\enddata
\end{deluxetable}


%
%








\clearpage

\begin{thebibliography}{}
\bibitem[Calzetti et al. 1994]{cal94} Calzetti, D., Kinney, A.L. \&
Storchi-Bergmann, T. 1994, \apjl, 429, 582
\bibitem[Cardelli et al. 1987]{car87} Cardelli, J. A., Clayton, G. C. \&
Mathis, J. S. 1987, \apj, 345, 245
\bibitem[Castander \& Lamb 1998]{CL98} Castander, F.J. \& Lamb, D.Q. 1998,
\apj, submitted, (CL98).
\bibitem[Costa et al. 1997a]{cos97a} Costa, E., Feroci, M., Frontera, F.,
Zavattini, G., Nicastro, L., Palazzi, E., Spoliti, G., Di Ciolo, L.,
Coletta, A., D'Andreta, G, et al, 1997a, IAU Circ 6572 
\bibitem[Costa et al. 1997b]{cos97b} Costa, E., Feroci, M., Piro, L., Cinti,
M.N., Frontera, F., Zavattini, G., Nicastro, L., Palazzi, E., Dal Fiume,
D., Orlandini, M., et al, 1997b, IAU Circ 6576 
\bibitem[Djorgovski et al. 1997]{djo97} Djorgovski, S.G., Kulkarni, S.R.,
Gal, R.R., Odewahn, S.C. \& Frail, D.A. 1997, IAU Circ 6732
\bibitem[Fioc \& Rocca-Volmerange 1997]{FRV97} Fioc, M. \&
Rocca-Volmerange, B., 1997, \aap, 326, 950.
\bibitem[Frontera et al. 1997]{fro97} Frontera, F., Greiner, J., Antonelli,
L.A., Dal Fiume, D., Orlandini, M., Boller, T, Woges, W., et al, 1997,
IAUCirc 6637
\bibitem[Frontera et al. 1998]{fro98} Frontera, F., Greiner, J., Antonelli,
L.A., Costa, E., Fiore, F., Parmar, A.N., Piro, L., Boller, T. \& Woges,
W., 1998b, \aap, in press, astro-ph/9804270.
\bibitem[Fruchter et al. 1997]{fru97} Fruchter, A., Livio, M., Macchetto,
D., Petro, L.,  Sahu, K., Pian, E., Frontera, F., Thorsett, S. \& Tavani,
M., 1997, IAU Circ 6747
\bibitem[Fruchter et al. 1998]{fru98} Fruchter, A., Pian, E., Thorsett, S.,
Gonz\'alez, R., Sahu, K., Mutchler, M., Frontera, F., Galama, T., Groot,
P., Hook, R., Kouveliotou, C., Livio, M., Macchetto, D., van Paradijs, J.,
Palazzi, E., Petro, L. \& Tavani, M., 1998, Proceedings of 4th Huntsville
Gamma-Ray Burst Symposium, eds. C. A. Meegan, R. Preece, and T. Koshut,
astro-ph/9801169
\bibitem[Galama et al. 1997]{gal97} Galama, T., Groot, P., van Paradijs,
J., Kouveliotou, C., Robinson, C.R., Fishman, G.J., Meegan, C.A., Sahu,
K.C., Livio, M., Petro, L., Macchetto, F.D., Heise, J., in't Zand, J.,
Strom, R.G., Telting, J., Rutten, R.G.M., Pettini, M., Tanvir, N. \& Bloom,
J.  1997, \nat, 387, 479
\bibitem[Galama et al. 1998]{gal98} Galama, T., Groot, P., van Paradijs, J.,
Kouveliotou, C., Sahu, K.C., Livio, M., Petro, L., Macchetto, F.D. \&
Fruchter, A.  1998, Proceedings of 4th Huntsville Gamma-Ray Burst
Symposium, eds. C. A. Meegan, R. Preece, and T. Koshut, astro-ph/9712322
\bibitem[Groot et al. 1997a]{gro97a} Groot, P., Galama, T., van Paradijs, J.,
Strom, R., Telting, J., Rutten, R.G.M., Pettini, M., Tanvir, N., et al,
1997a, IAU Circ 6584
\bibitem[Groot et al. 1997b]{gro97b} Groot, P., Galama, T., van Paradijs, J.,
Melnick, G., van der Steene, G., Bremer, M., Tanvir, N., Bloom, J., Strom,
R., Telting, J., Rutten, R.G.M., et al, 1997b, IAU Circ 6588
\bibitem[Guarnieri et al. 1997]{gua97} Guarnieri, A., Bartolini, C.,
Masetti, N., Piccioni, A., Costa, E., Feroci, M., Frontera, F., Dal Fiume,
D., Nicastro, L., Palazzi, E., Castro-Tirado, A.J. \& Gorosabel, J., 1997,
\aap, 328, L13
\bibitem[Holtzman et al. 1995a]{hol95a} Holtzman, J., Hester, J.J.,
Casertano, S., Trauger, J.T., Watson, A.M., Ballester, G.E., Burrows, C.J.,
Clarke, J.T. et al, 1995a, \pasp, 107, 156
\bibitem[Holtzman et al. 1995b]{hol95b} Holtzman, J., Burrows, C.J.,
Casertano, S., Hester, J.J., Trauger, J.T., Watson, A.M. \& Worthey,
G. 1995b, \pasp, 107, 1065
\bibitem[Klose et al. 1997]{klo97} Klose, S., Stecklum, B. \& Tuffs,
R. 1997, IAU Circ 6611
\bibitem[Kulkarni et al. 1997]{kul97} Kulkarni, S.R., Djorgovski, S.G. \&
Clemens, 1997, IAU Circ 6732
\bibitem[Livio et al. 1998]{liv98} Livio, M., Sahu, K., Petro, L., Fruchter,
A., Pian, E., Macchetto, D., van Paradijs, J., Kouveliotou, C., Groot,
P. \& Galama, T., 1998, Proceedings of 4th Huntsville Gamma-Ray Burst
Symposium, eds. C. A. Meegan, R. Preece, and T. Koshut, astro-ph/9712097
\bibitem[Margon et al. 1997]{mar97} Margon, B., Deutch, E.W., Lamb, D.Q. \&
Castander, F.J. 1997, IAU Circ 6618
\bibitem[Metzger et al. 1997a]{met97a} Metzger, M.R., Kulkarni, S.R.,
Djorgovski, S.G., Gal, R., Steidel, C.C., 1997a, IAU Circ 6588
\bibitem[Metzger et al. 1997b]{met97b} Metzger, M.R., Cohen, J.L.,
Blaceslee, J.P., Kulkarni, S.R., Djorgovski, S.G., Steidel, C.C. \& Frail,
D.A. 1997b, IAU Circ 6631
\bibitem[O'Donnell 1994]{odo94} O'Donnell, J.E. 1994, \apj, 422, 158
\bibitem[Pedichini et al. 1997a]{ped97a} Pedichini, F., Di Paola, A., Stella,
L., Gandolfi, G., Spoliti, G., Di Ciolo, L., Coletta, A., D'Andreta, G., Muller, J.M., Capalbi, M., Boattini, A.,  Costa, E., Feroci, M., Piro,
L., Palazzi, E., Dal Fiume, D., Nicastro, L., Frontera, F., Heise, J. \& In't
Zand, J. 1997, IAU Circ 6635
\bibitem[Pedichini et al. 1997b]{ped97b} Pedichini, F., Di Paola, A., Stella,
L., Buonanno, R., Boattini, A., Gandolfi, G., Costa, E., Feroci, M., Piro,
L., Dal Fiume, D., Frontera, F., Nicastro, L., Palazzi, E., Heise, J., In't
Zand, J. \& Vietri, M., 1997, \aap, 327, L32
\bibitem[Reichart 1997]{rei97} Reichart, D.E. 1998, \apjl, 485, L57
\bibitem[Robison 1997]{rob97} Robinson, R.D. 1997, ``Examining the STIS
point spread function'',
http://www.stsci.edu/ftp/instrument\_news/STIS/performance/psf/psf\_robinson.ps
\bibitem[Sahu et al. 1997a]{sah97a} Sahu, K. C., Livio, M., Petro, L. \&
Macchetto, F., 1997a, IAU Circ 6606
\bibitem[Sahu et al. 1997b]{sah97b} Sahu, K. C., Livio, M., Petro, L.,
Macchetto, F., van Paradijs, J., Kouveliotou, C., Fishman, G. \& Meegan,
C., 1997b, IAU Circ 6619
\bibitem[Sahu et al. 1997c]{sah97c} Sahu, K. C., Livio, M., Petro, L.,
Macchetto, F. D., Van Paradijs, J., Kouveliotou, C., Fishman, G. J., Meegan,
C. A., Groot, P. J. \& Galama, T. 1997, \nat, 387, 476
\bibitem[Soifer et al. 1997]{soi97} Soifer, B., Neugebauer, G., Armus, L.,
Metzger, M., Kulkarni, S., Djorgovski, S., Steidel, C. \& Frail, D. 1997,
IAU Circ 6619
\bibitem[Tonry et al. 1997]{ton97} Tonry, J. L., Hu, E. M., Cowie, L. L. \&
McMahon, R. G. 1997, IAUCirc 6620 
\bibitem[van Paradijs et al. 1997]{vpa97} van Paradijs, J., Groot, P.,
Galama, T., Kouveliotou, C., Strom, R.G., Telting, J., Rutten, R.G.M.,
Fishman, G.J., Meegan, C.A., Pettini, M., Tanvir, N., et al. 1997, \nat,
386, 686
\bibitem[Wijers et al. 1997]{wij97} Wijers, R.M.A.J., Rees, M.J. \& M\'esz\'aros, P. 1997, \mnras, 288, L51
\bibitem[Yoshida et al. 1997]{yos97} Yoshida, A., Kawai, N., Otani, C.,
Tokanai., F., Inoue, H., Murakami, T., Nagase, F., Shibata, R., et al
1997, IAUCirc 6593  
\end{thebibliography}
\end{document}